 \documentclass[aps,prl,twocolumn,showpacs]{revtex4-1}
 
\usepackage[]{graphicx}
\usepackage{amsmath}
\usepackage{times}
\usepackage{bm}
\usepackage{units}
\usepackage{ulem}
\usepackage{color}

\newcommand{\comment}[1]{}

\begin{document}

\title{Microscopic mechanism for transient population inversion and optical gain in graphene}

\author{Torben Winzer}
\email[]{t.winzer@mailbox.tu-berlin.de}
\author{Ermin Mali\'{c}}
\author{Andreas Knorr}

\affiliation{Institut f\"ur Theoretische Physik, Nichtlineare Optik und Quantenelektronik, Technische Universit\"at Berlin,
  Hardenbergstr. 36, 10623 Berlin, Germany}

\begin{abstract}
 A transient femtosecond population inversion in graphene was recently reported by Li \textit{et al.}, Phys. Rev. Lett. \textbf{108}, 167401 (2012). Based on a microscopic theory we clarify the underlying microscopic mechanism: Transient gain and population inversion in graphene occurs due to a complex interplay of strong optical pumping and carrier cooling that fills states close to the Dirac point giving rise to a relaxation bottleneck. The subsequent femtosecond decay of the optical gain is mainly driven by Coulomb-induced Auger recombination. 
\end{abstract}

\maketitle

Light amplification and optical gain are of central importance in the optical sciences \cite{haken84,scully97}. They are based on the quantum effect of induced photon emission, and constitute one of the key ingredients in optoelectronic devices and optical data communication \cite{scully97}. Therefore, there is an ongoing quest for gain materials in different frequency ranges. Particular questions involve the mechanical stability of materials and the possibility of electrical injection. In recent years much research is devoted to graphene, a two dimensional sheet of carbon atoms \cite{geim07,castro_neto09,bonaccorso10,avouris12}. Beside its distinct transport characteristics \cite{Zhang05}, its optical properties have been intensively investigated  \cite{bonaccorso10,sun10-1,avouris12} with respect to transient broadband \cite{mak11,Mak12} and saturable \cite{Bao09,sun10} absorption.

Recently, T. Li and coworkers \cite{Li12} demonstrated also the possibility of optical gain based on population inversion in optically pumped graphene within a differential reflection measurement. The experiment reveals that gain appears at pump fluences of approximately $\unit[2]{mJ/cm^2}$ on a femtosecond timescale for optical transitions smaller than the pump energy \cite{Li12}. Graphene as a coherent source of $T\!H\!z$ radiation was first suggested by Ryzhii {\textit{et al.}} \cite{ryzhii07,satou08} and experimental indications were recently found in an optical non-differential measurement \cite{tombet12}. Since graphene has no band gap and its optical properties are dominated by distinct many body effects, the observation of optical gain in graphene is crucial to our understanding of the light-matter interaction in this material. In particular, it is of outermost importance for graphene-based applications to understand the conditions for the occurrence of optical gain.

 \begin{figure}[t!]
  \begin{center}
\includegraphics[width=\linewidth]{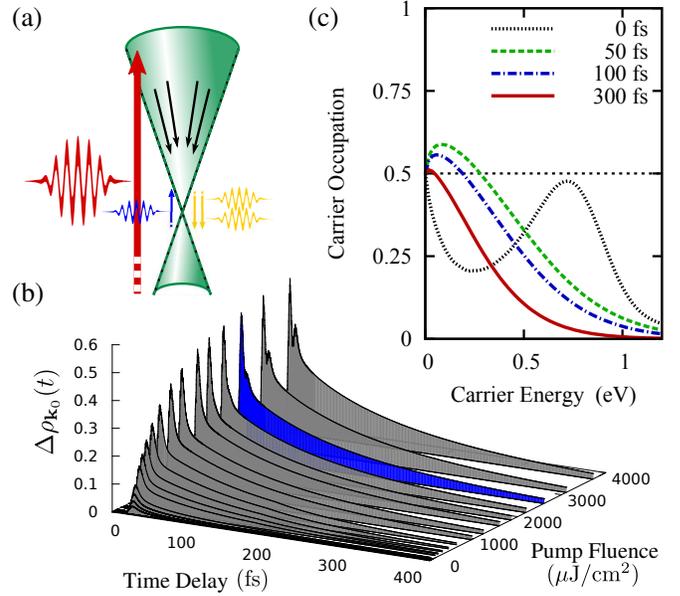}
  \end{center}
  \caption{(a) Scheme of optical pumping (red) and the build-up of a phonon bottleneck (black). Energetically lower probe pulse (blue) can be amplified via induced emission (yellow), if population inversion occurs.  (b) Pump-induced change of occupation $\Delta\rho_{{\bf k}_0}(t)$ in the most efficiently excited state ${\bf k}_0$ for different pump fluences. We obtain a qualitatively and quantitatively excellent agreement with the experiment, cf. Fig. 2(b) in Ref. \cite{Li12}. (c) In contrast to the experiment \cite{Li12}, we have a direct access to the carrier dynamics: Carrier distribution for different delay times after an optical excitation with a pump fluence of approximately $\unit[2.5]{mJ/cm^2}$ (marked blue in (b)). Gain is obtained during the first $\unit[300]{fs}$ for carrier energies smaller than $\unit[250]{meV}$.}\label{FIG1}  
\end{figure} 

 In this Letter, we perform microscopic calculations on the carrier dynamics in the same excitation regime, as recently reported in experiments \cite{Li12}, confirming the possibility of gain and population inversion in graphene. We develop a complementary understanding with respect to the experiment, since our theory directly accesses the time- and momentum-resolved non-equilibrium carrier distribution during and after a strong optical excitation pulse. Therewith, we are able to analyze the predominant interplay of many body mechanisms (electron-phonon and electron-electron interaction) and light-matter interaction giving rise to the gain regime. In particular, as the most important result of this paper, we identify as the crucial process the phonon-induced intraband scattering, which fills optically active states at low energies. This process is responsible for a carrier relaxation bottleneck close to the Dirac point resulting in a population inversion. The temporally subsequent decay of the population inversion on a hundred femtosecond timescale can be understood as an efficient interplay of Coulomb-induced Auger recombination \cite{rana07,winzer10} (carrier loss) and intraband carrier-phonon scattering (energy loss). Signatures of these collinear scattering channels \cite{rana07}  were recently also observed in photo-current experiments \cite{gabor11,sun12} and optical pump-probe measurements \cite{obraztsov11,tani12,*tani12-1}.

Applying the graphene Bloch-equations \cite{Malic11,Stroucken12}, we microscopically consider the light-carrier interaction as well as carrier-carrier and carrier-phonon scattering  in a consistent treatment \cite{lindberg88, Rossi02}. A coupled set of differential equations describes the temporal evolution of (i) the occupation probability $\rho^{\lambda}_{\bf k}$ in the state $\bf k$ in the conduction and the valence band ($\lambda=c,v$), (ii) the microscopic polarization $p_{\bf k}$ reflecting the transition probability between both bands,  and (iii) the phonon occupation $n^j_{\bf q}$ with the momentum $\bf q$ for different modes $j$ \cite{Malic11}:
\begin{equation}\label{GBE}
 \begin{split}
  \dot{\rho}_{\bf k}^{\lambda}&=2\Im\left[\Omega_{\bf k}^{vc*}p_{\bf k}\right]+\Gamma_{{\bf k},\lambda}^{in}\left[1-\rho_{\bf k}^{\lambda}\right]-\Gamma_{{\bf k},\lambda}^{out}\rho_{\bf k}^{\lambda},\\
\dot{p}_{\bf k}&=\left[i\Delta\omega_{\bf k}-i\Omega_{\bf k}^{\lambda\lambda}-\gamma_{\bf k}\right]p_{\bf k}-i\Omega_{\bf k}^{vc}\left[\rho_{\bf k}^c-\rho_{\bf k}^v\right]+\mathcal{U}_{\bf k},\\
\dot{n}_{\bf q}^j&=-\gamma_{ph}\left[n_{\bf q}^j-n_B^j\right]+\Gamma_{j,{\bf q}}^{em}\left[1+n_{\bf q}^j\right]-\Gamma_{j,{\bf q}}^{abs}n_{\bf q}^j.
 \end{split}
\end{equation}

The coupling to the exciting pulse is described by the Rabi frequency $\Omega_{\bf k}^{vc}$ and its corresponding intraband contribution $\Omega_{\bf k}^{\lambda\lambda}$. For our analysis we generate a non-equilibrium carrier distribution via optical excitation, similar to the experimental realization \cite{Li12}: a laser pulse in $\Omega_{\bf k}^{\lambda\lambda'}$ with a duration of $\unit[10]{fs}$ and a photon energy of $\unit[1.5]{eV}$ is applied. A sketch of excitation and the linear dispersion is shown in Fig. \ref{FIG1}(a). A delayed probe pulse (blue) is used to test the pump (red) induced population change. Many particle interactions ($\Gamma$, $\mathcal{U}_{\bf k}$, and $\gamma_{\bf k}$) are treated within second-order Born-Markov approximation, where we take into account a self-consistently determined many-particle broadening of the strict energy conservation \cite{schilp94}. We obtain a Boltzmann-like scattering equation, where the time- and momentum-dependent scattering rates $\Gamma_{{\bf k},\lambda}^{in/out}$ consider carrier-carrier as well as carrier-phonon contributions. A similar equation is obtained for the phonon occupation with the emission and absorption rates $\Gamma_{j,{\bf q}}^{em/abs}$. The microscopic polarization $p_{\bf k}$ is damped by the many-particle diagonal dephasing $\gamma_{\bf k}$ and deformed by the off-diagonal contribution $\mathcal{U}_{\bf k}$ \cite{Malic11}. In exception of the linear transition frequency $\Delta\omega_{\bf k}$, the experimentally estimated phonon decay rate $\gamma_{ph}$ \cite{Kang10}, and the Bose-Einstein distribution $n_B^j$ as a thermal bath for the dynamic phonon occupation, all terms in Eqs. (\ref{GBE}) are explicitly time-dependent. A detailed description of these equations can be found in Ref. \cite{Malic11}.

For the understanding of the formation of optical gain we analyze as an experimentally addressable observable the pump pulse-induced population inversion $\Delta \rho_{\bf k}(t)$ at the position of the probe field $|{\bf k}_0|=\omega_L/2v_F$, where $\omega_L$ is the probe frequency and $v_F$ is the carrier velocity in graphene's linear band structure. From the numerical solution of the graphene Bloch equations (\ref{GBE}) we obtain $\Delta \rho_{{\bf k}_0}(t)$, which is approximately proportional to the differential optical response, e.g. the reflection $\Delta R/R$ \cite{Breusing11,winnerl11}.

Figure \ref{FIG1}(b) shows $\Delta \rho_{{\bf k}_0}(t)$ for different pump fluences up to $\unit[4]{mJ/cm^2}$. In excellent agreement with the experiment \cite{Li12}, we observe in  the low excitation regime a linear scaling of the peak of  $\Delta \rho_{{\bf k}_0}(t)$ with the pump fluence, and a clearly nonlinear saturation effect in the range of $\unit[0.5-1]{\mu J/cm^2}$. At very high fluences, we find first signatures of Rabi oscillations. The gain (population inversion) regime can be directly observed in the energy-resolved carrier distribution, shown in Fig. \ref{FIG1}(c) for an exemplary pump fluence of $\unit[2.5]{\mu J/cm^2}$ (marked blue in Fig. \ref{FIG1}(b)) at different times after the optical excitation. At time $t=0$, i.e. at the maximum of the excitation pulse, a peaked non-equilibrium population distribution is found. The low energy occupation wing stems from the initial thermal distribution at room temperature broadened by relaxation processes already acting during the pulse. In the course of time, after the pulse, the low energy population increases and we observe a population inversion for carrier energies of up to $\unit[250]{meV}$. An analysis of the distribution shows that as long as the population inversion is present no Fermi-Dirac distribution is established. A non-equilibrium between the carrier and phonon system is present. Depending on the pump fluence, it takes a few hundreds of $\unit[]{fs}$ until equilibrium is reached.

\begin{figure}[t!]
  \begin{center}
\includegraphics[width=0.87\linewidth]{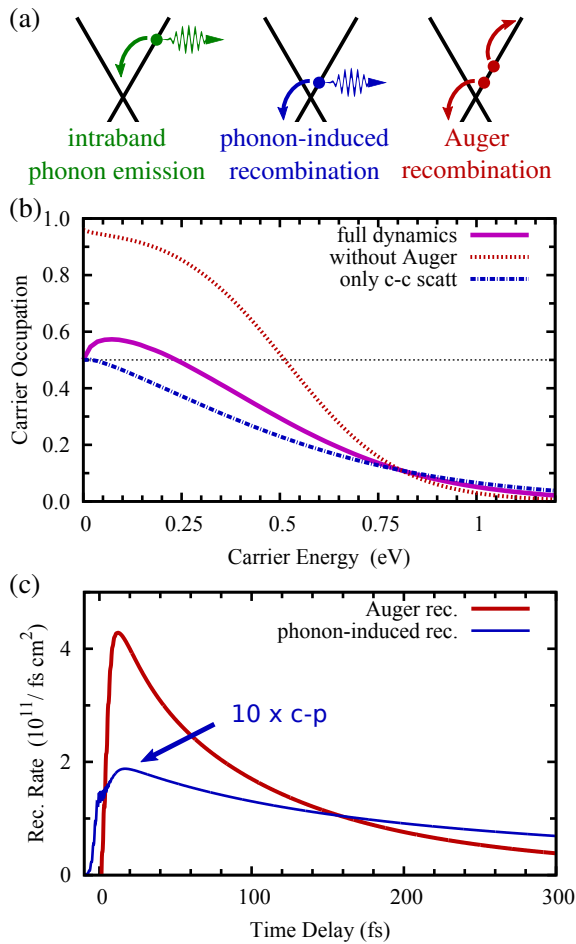}
  \end{center}
  \caption{(a) Sketch of the most important scattering channels: energy loss due to intraband phonon emission (green, left), phonon-induced recombination (blue, middle), and Coulomb-induced Auger recombination (red). (b) Carrier distribution $\unit[50]{fs}$ after the excitation pulse ($\unit[2.5]{\mu J/cm^2}$) considering the full dynamics (purple line), only Coulomb-induced scattering processes (dashed-dotted blue line), and all relaxation channels in exception of Auger-processes (dotted red line). The figure illustrates that the intraband carrier-phonon scattering is responsible for the occurrence of the population inversion, while Auger processes are counteracting. (c) Coulomb-  (red) and phonon-induced (blue) recombination rates, reflecting their contribution to the decay of the gain regime, where the process of Auger recombination is the predominant mechanism.} \label{FIG2} 
\end{figure} 

Next we turn to the microscopic explanation of the formation mechanism of the population inversion. A detailed inspection shows, that the population inversion at lower energies is a direct consequence of energy loss of the optically excited carriers due to intraband carrier-phonon scattering, cf. Fig. \ref{FIG2}(a) (green, left). This process fills the states with low momenta having a low (linear) density of states. Hereby, the interplay of the reduced density of states (vanishing at $|k|=0$) and the conserved carrier density results in a higher occupation at lower energies in comparison to the optically excited distribution, where no population inversion is found.  This process of energy loss within a band is indirectly accelerated by an ultrafast broadening of the non-equilibrium carrier distribution via intraband Coulomb-scattering. A counteracting process is phonon-induced recombination between the bands, cf. Fig. \ref{FIG2}(a) (blue, middle). However, in graphene it is less efficient due to the reduced density of states close to the Dirac point, resulting in a phonon bottleneck \cite{Butscher07}.

 This overall picture of the population inversion mechanism is supported by Fig. \ref{FIG2}. The population distribution $\unit[50]{fs}$ after the excitation pulse under consideration of different relaxation channels is shown in Fig. \ref{FIG2}(b). For the purple line all scattering channels are considered and a population inversion is obtained for carrier energies smaller than $\unit[250]{meV}$. Next, we neglect carrier-phonon scattering to demonstrate its crucial role for the build-up of a population inversion (blue dashed-dotted line in Fig. \ref{FIG2}(b)). Without this mechanism of intraband energy loss, the population inversion is suppressed by efficient Coulomb-induced Auger recombination \cite{plochocka09,kim11,Winzer12}, sketched in Fig. \ref{FIG2}(a) (red). Its redistribution of carriers to higher energies results in a single Fermi distribution for conduction and valence band with a vanishing chemical potential \cite{Winzer12}, cf. Fig. \ref{FIG2}(b). Taking into account all relaxation channels but the Auger processes (red dotted line) gives rise to an energetically broad gain regime with an unrealistically extensive persistence. This reflects the role of Auger-induced recombination to the transience of the population inversion.

One direct consequence of the carrier recombination (Auger and electron-phonon) in the process of transient population inversion, which is in full agreement with experimental measurements \cite{Li12}, is a decay of the population inversion on a femtosecond timescale. Such a decay has been measured in the pump-probe signal in Ref. \cite{Li12}. To study the decay of the population inversion in more detail, we analyze the phonon- and Coulomb-induced recombination rates, shown in Fig. \ref{FIG2}(c). Our calculations reveal that in the strong excitation regime Coulomb-induced Auger processes (red line) are predominant recombination channels, whereas interband carrier-phonon scattering (blue line) plays a minor. In particular, immediately after the excitation pulse the Coulomb-induced Auger recombination reaches values of up to $\unit[4\cdot10^{11}]{/fs\hspace{0.5mm}cm^2}$, which are large compared to the phonon-induced recombination of only $\unit[2\cdot10^{10}]{/fs\hspace{0.5mm}cm^2}$ and the overall carrier density of $\unit[4\cdot10^{13}]{/cm^2}$. Consequently, the Auger recombination outweighs the phonon-induced intraband filling of the low-energy states resulting in a femtosecond decay of the population inversion.

 \begin{figure}[t!]
  \begin{center}
\includegraphics[width=\linewidth]{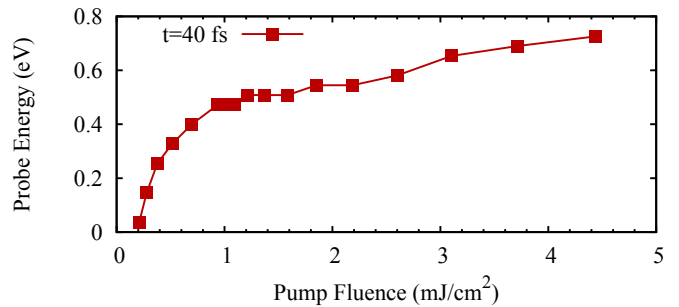}
  \end{center}
  \caption{Maximal optical probe energy, where a population inversion can be obtained depending on the pump fluence. The values are extracted from the carrier distribution $\unit[40]{fs}$ after the excitation pulse, corresponding to the temporal resolution of the experiment in Ref. \cite{Li12}.} \label{FIG3} 
\end{figure} 

A key quest concerning applications of graphene in optoelectronics is the frequency regime where optical gain can be obtained and its dependence on the excitation fluence. As reported in Ref. \cite{Li12}, for a certain excitation strength the carrier occupation at a probe energy of $\unit[1.55]{eV}$ remains beneath the gain regime, whereas at lower probe energies a population inversion was detected. We use our access to the momentum-resolved carrier distribution to determine the maximal optical probe energy, at which population inversion is obtained depending on the pump fluence, shown in Fig. \ref{FIG3}. The values are taken $\unit[40]{fs}$ after the excitation  pulse, corresponding to the temporal resolution of the experiment \cite{Li12}. Up to a critical pump fluence of approximately $\unit[200]{\mu J/cm^2}$ we observe no population inversion. Above this value the population inversion appears abruptly at low probe energies. For stronger excitations the gain regime is broadened up to $\unit[0.75]{eV}$. The experimentally observed population inversion at a probe energy of  $\unit[1.16]{eV}$ can be obtained for pump fluences $>\unit[5]{mJ/cm^2}$.

 \begin{figure}[t!]
  \begin{center}
\includegraphics[width=\linewidth]{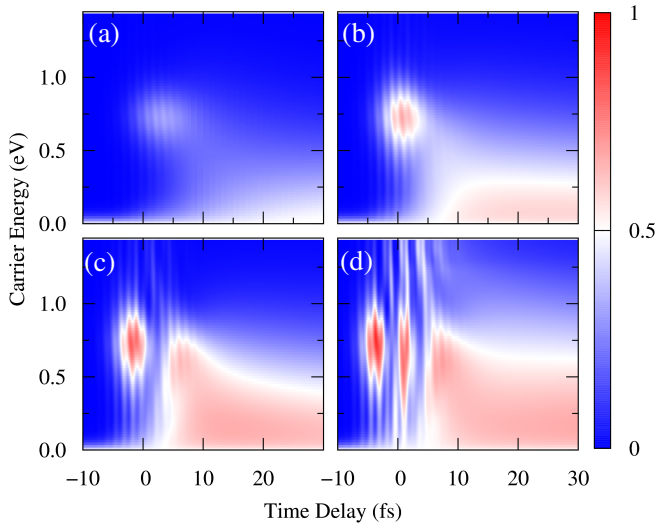}
  \end{center}
  \caption{ Time- and energy-resolved carrier distribution along the direction of the strongest light-matter coupling during excitations with different pump fluences: (a)  moderate excitation regime, (b) gain regime, (c) Rabi-oscillation regime, and (d) second-harmonic regime.} \label{FIG4} 
\end{figure} 

Finally, we analyze the energy-resolved carrier distribution during and after the optical excitation with an increasing pump fluence, cf. Fig. \ref{FIG4}(a-d). The shown occupations along the direction of the strongest light-matter coupling in the two-dimensional reciprocal space \cite{Malic11} directly illustrate the action of the exciting laser pulse \footnote{The results shown in Figs. \ref{FIG1}a,b and Fig. \ref{FIG2}b stemming from angle-averaged carrier distributions.}. The rise of the non-equilibrium occurs wave-like reflecting the optical pump frequency. In exception of panel (a), where a moderate excitation strength beneath the critical value was used, we observe a gain regime (red area) in a broad energy range depending on the applied pump fluence. Note that we also obtain population inversion at the excitation energy decaying on the time scale of the pump pulse duration. However, this should not be visible in experiments using the same temporal resolution for pump and probe pulse. Furthermore, the energy and time resolution of the carrier distribution uncovers different aspects of the light-matter coupling in strong excitation regime. The excitation used in Fig. \ref{FIG4}(c) gives rise to Rabi oscillations, since optical pumping overweights the diagonal dephasing $\gamma_{\bf k}$ for the microscopic polarization $p_{\bf k}$ , cf. Eqs. (\ref{GBE}). For the highest pump fluence, shown in Fig. \ref{FIG4}(d), even a wing of the coherent two-photon absorption is visible.

In conclusion, our calculations confirm the possibility of transient optical gain and population inversion in optically pumped graphene. We identify intraband energy loss via optical phonons as the predominant mechanism for the occurrence of the population inversion. Its decay on a femtosecond time scale is mainly driven by the process of Auger recombination.

     
We acknowledge the financial support from the Deutsche Forschungsgemeinschaft
through SPP 1459. E. M. thanks the Einstein Foundation Berlin.

%

\end{document}